# DendroPy 5: a mature Python library for phylogenetic computing

Matthew Andres Moreno  [1,2,3], Mark T. Holder  [4,5], and Jeet Sukumaran  [6]

**1** Department of Ecology and Evolutionary Biology, University of Michigan, Ann Arbor, MI, USA **2** Center for the Study of Complex Systems, University of Michigan, Ann Arbor, MI, USA **3** Michigan Institute for Data Science, University of Michigan, Ann Arbor, MI, USA **4** Department of Ecology and Evolutionary Biology, University of Kansas, Lawrence, KS, USA **5** Biodiversity Institute, University of Kansas, Lawrence, KS, USA **6** Department of Biology, San Diego State University, San Diego, CA, USA

## Summary

Contemporary bioinformatics has seen in profound new visibility into the composition, structure, and history of the natural world around us. Arguably, the central pillar of bioinformatics is phylogenetics — the study of hereditary relatedness among organisms. Insight from phylogenetic analysis has touched nearly every corner of biology. Examples range across natural history  Title et al., 2024), population genetics and phylogeography (Knowles & Maddison, 2002), conservation biology (Faith, 1992), public health (Giardina et al., 2017; Voznica et al., 2022), medicine (Kim et al., 2006; Lewinsohn et al., 2023), *in vivo* and *in silico* experimental evolution Moreno et al. (2023), application-oriented evolutionary algorithms (Hernandez et al., 2022; Lalejini et al., 2024; Shahbandegan et al., 2022), and beyond.

High-throughput genetic and phenotypic data has realized groundbreaking results, in large part, through conjunction with open-source software used to process and analyze it. Indeed, the preceding decades have ushered in a flourishing ecosystem of bioinformatics software applications and libraries. Over the course of its nearly fifteen-year history, the DendroPy library for phylogenetic computation in Python has established a generalist niche in serving the bioinformatics community (Sukumaran & Holder, 2010). Here, we report on the recent major release of the library, DendroPy version 5. The software release represents a major milestone in transitioning the library to a sustainable long-term development and maintenance trajectory. As such, this work positions DendroPy to continue fulfilling a key supporting role in phyloinformatics infrastructure.

## Statement of Need

DendroPy operates within a rich ecosystem of packages, frameworks, toolkits, and software projects supporting bioinformatics and phylogenetics research. The broader software landscape largely divides into the following major categories,

1. High-performance specialized tools for inference (e.g., *BEAST2*, *RAxML*, *MrBayes*, *PAUP*, etc.) (Bouckaert, 2014; Ronquist et al., 2012; Stamatakis, 2014; Wilgenbusch & Swofford, 2003);
2. Python phylogenetics libraries that provide rich tree-centric data models and operations, such as
   - *ETE*, known in particular for powerful phylogeny visualization capabilities (Huerta-Cepas et al., 2016),
   - *Scikit-bio* and tskit (Jai Ram Rideout et al., 2024; Kelleher et al., 2018),

- *TreeSwift* and *SuchTree*, which provide lightweight, high-performance tree representations (Moshiri, 2020; Y. Neches & Scott, 2018), and
- *hstrat* and *Phylotrack*, which specialize in collecting phylogenies from agent-based evolutionary simulation (Dolson et al., 2024; Moreno et al., 2022);

3. Python phylogenetics libraries with genome/gene-centric data models and operations (e.g., *PyCogent*/*Cogent3*, *BioPython*, etc.) (Cock et al., 2009; Knight et al., 2007); and
4. numerous R phylogenetics packages, which are often highly specialized but generally interoperate via `ape.phylo` data structures (Paradis & Schliep, 2019).

DendroPy falls largely within the second camp above. It is notable in providing a broad portfolio of evolutionary models, but also fields population genetics and sequence evolution utilities. DendroPy is also notable for its comprehensive, systematic documentation and rich, user-extensible tree representation. The library's use cases range across serving as a stand-alone library for phylogenetics, a component of more complex multi-library phyloinformatics pipelines, or as an interstitial "glue" that assembles and drives such pipelines.

## Features

Key features of DendroPy include,

- rich object-oriented representations for manipulation of phylogenetic trees and character matrices,
- efficient, bit-level representation of nodes' leaf bipartitions,
- loading and saving popular phylogenetic data formats, including NEXUS, Newick, NeXML, Phylip, and FASTA (Felsenstein, 1981; Lipman & Pearson, 1985; Maddison et al., 1997; Olsen, 1990; Vos et al., 2012);
- simulation of phylogenetic trees under a range of models, including coalescent models, birth-death models, and population genetics simulations of gene trees; and
- application scripts for performing data conversion, collating taxon labels from multiple trees, and tree posterior distribution summarization.

## Maintenance

A major focus of DendroPy's version 5 release has been in establishing a trajectory for sustainable long-term maintenance. Our intention is that effort reducing maintenance burden will translate into a regular release schedule (incorporating timely patches for reported issues), development of new features, and incorporation of user contributions.

The version 5 release reflects substantial investment in adopting modern software development best practices. In version 5, DendroPy has officially dropped support for Python 2.7, as well as Python 3.X versions that have reached end-of-life. Focusing support on Python 3.6 and higher simplifies cross-environment testing and allows future development to leverage new language features. In addition, we have established comprehensive continuous integration (CI) infrastructure via GitHub Actions, comprising

- code linting with Ruff
- deploying up-to-date documentation via GitHub pages,[1]
- unit tests, largely organized within the `unittest` framework,
- new smoke tests using pytest,
- code coverage reporting via the Codecov service, and
- automatic deployment of tagged versions to PyPI.

Other behind-the-scenes activities in preparing this release included repair of library components flagged by the new tooling, triage of user bug reports, applying issue tags to manage open tracker items, establishing a code of conduct, and creating issue templates to increase the quality of future bug reports and feature requests. Altogether, these improvements serve as a foundation for future work maintaining and extending DendroPy in a manner that is reliable, stable, and responsive to user needs. We look forward to this next chapter.

---

[1] Documentation is hosted at https://jeetsukumaran.github.io/DendroPy.

## Impact

Over its nearly 15-year history, DendroPy's versatility and stability have driven adoption as a core dependency of many phylogenetics pipelines and bioinformatics software libraries. Currently, [85 projects](#) on PyPI list DendroPy as a direct dependency. Notable projects using DendroPy include:

> PASTA, which performs multiple sequence alignment ([Mirarab et al., 2014](#));
> Physcraper, which automates curation of gene trees ([Sánchez-Reyes et al., 2021](#));
> Propinquity, the supertree pipeline ([Redelings & Holder, 2017](#)) of the Open Tree of Life project;
> DELINEATE, software for analyses discerning true speciation from population lineages ([Sukumaran et al., 2021](#));
> Archipelago, which models spatially explicit biographical phylogenesis ([Sukumaran et al., 2015](#));
> Espalier, a utility for constructing maximum agreement forests ([Rasmussen & Guo, 2023](#)); and
> MetaPhlAn, which extracts information about microbial community composition from metagenomic shotgun sequencing data ([Blanco-Míguez et al., 2023](#)).

During this time, DendroPy has also directly helped enable numerous end-user phylogenetics projects. Notable examples include work on the early natural history of birds ([Jarvis et al., 2014](#)), the molecular evolution of the Zika virus ([Faye et al., 2014](#)), and early human migration within the Americas ([García-Ortiz et al., 2021](#)). As of May 2024, Google Scholar counts 1,654 works referencing DendroPy ([Sukumaran & Holder, 2010](#)).

## Acknowledgements

Thank you to University of Michigan Undergraduate Research Opportunity Program participant Connor Yang for his contributions in increasing test coverage, and to our open-source community for bug reports, feature suggestions, and patch contributions over the years. This research is based upon work supported by:

> the Eric and Wendy Schmidt AI in Science Postdoctoral Fellowship, a Schmidt Futures program (author MAM)
> the National Science Foundation grant NSF-DEB 1937725 "COLLABORATIVE RESEARCH: Phylogenomics, spatial phylogenetics and conservation prioritization in trapdoor spiders (and kin) of the California Floristic Province" (author JS)
> the National Science Foundation grant NSF-DEB 1457776 "Collaborative Research Developing novel methods for estimating coevolutionary processes using tapeworms and their shark and ray hosts" (author MH)